\documentclass[12pt]{article}
\usepackage{epsf}
\def\beq{\begin{equation}}
\def\eeq{\end{equation}}
\def\drcn{\delta R^{c/n}}
\def\ap#1#2#3 {Ann. Phys. (NY) {\bf#1} (19#2) #3}
\def\err#1#2#3 {{\it Erratum} {\bf#1} (19#2) #3}
\def\ib#1#2#3 {{\it ibid.} {\bf#1} (19#2) #3}
\def\ijmp#1#2#3 {Int. J. Mod. Phys. {\bf#1} (19#2) #3}
\def\jetp#1#2#3 {JETP Lett. {\bf#1} (19#2) #3}
\def\mpl#1#2#3 {Mod. Phys. Lett. {\bf#1} (19#2) #3}
\def\np#1#2#3 {Nucl. Phys. {\bf#1} (19#2) #3}
\def\pl#1#2#3 {Phys. Lett. {\bf#1} (19#2) #3}
\def\prep#1#2#3 {Phys. Rep. {\bf#1} (19#2) #3}
\def\prev#1#2#3 {Phys. Rev. {\bf#1} (19#2) #3}
\def\prl#1#2#3 {Phys. Rev. Lett. {\bf#1} (19#2) #3}
\def\sjnp#1#2#3 {Sov. J. Nucl. Phys. {\bf#1} (19#2) #3}
\def\spj#1#2#3 {Sov. Phys. JETP {\bf#1} (19#2) #3}
\def\spu#1#2#3 {Sov. Phys. Usp. {\bf#1} (19#2) #3}
\def\zp#1#2#3 {Zeit. Phys. {\bf#1} (19#2) #3}

\begin{document}
\begin{titlepage}
\begin{center}
{\Large \bf William I. Fine Theoretical Physics Institute \\
University of Minnesota \\}  \end{center}
\begin{flushright}
FTPI-MINN-07/08 \\
UMN-TH-2541/07 \\
LPT-Orsay/07-19\\
March 2007 \\
\end{flushright}
\begin{center}
{\Large \bf  Isospin breaking in the yield of heavy meson pairs in $e^+e^-$
annihilation near threshold
\\}
\vspace{0.1in}
{\bf S. Dubynskiy}$^a$, {\bf A. Le Yaouanc}$^b$, {\bf L. Oliver}$^b$, {\bf
J.-C.
Raynal}$^b$ {\bf and M.B. Voloshin}$^c$  \\[0.1in]
{\it $^a$ School of Physics and Astronomy, University of Minnesota, Minneapolis,
MN
55455, USA\\
$^b$ Laboratoire de Physique Th\'eorique\footnote{Unit\'e Mixte de
Recherche UMR
8627 - CNRS }, Universit\'e de Paris XI,
B\^atiment 210, 91405 Orsay Cedex,
France \\
$^c$ William I. Fine Theoretical Physics Institute, University of
Minnesota, Minneapolis, MN 55455, USA
and
Institute of Theoretical and Experimental Physics, Moscow, 117259, Russia}
\\[0.1in]
\end{center}

\begin{abstract}
We revisit the problem of interplay between the strong and the Coulomb
interaction in the charged-to-neutral yield ratio for $B {\bar B}$ and $D {\bar
D}$ pairs near their respective thresholds in $e^+e^-$ annihilation. We consider
here a realistic situation with a resonant interaction in the isospin $I=0$
channel and a nonresonant strong scattering amplitude in the $I=1$ state. We
find that the yield ratio has a smooth behavior depending on the scattering
phase in the $I=1$ channel. The same approach is also applicable to the $K {\bar
K}$ production at the $\phi(1020)$ resonance, where the Coulomb effect in the
charged-to-neutral yield ratio is generally sensitive to the scattering phases
in both the isoscalar and the isovector channels. Furthermore, we apply the same
approach to the treatment of the effect of the isotopic mass difference between
the charged and neutral mesons and argue that the strong-scattering effects
generally result in a modification to the pure kinematical effect of this
mass difference.
\end{abstract}
\end{titlepage}

\section{Introduction}
The $J^{PC}=1^{--}$ resonances near the new flavor thresholds: $\Upsilon(4S)$,
$\psi(3770)$, and $\phi(1020)$ are the well known sources in $e^+e^-$
experiments of pairs of the new-flavor mesons: respectively $B {\bar B}$, $D
{\bar D}$, and $K {\bar K}$. A number of experimental approaches depends on the
knowledge of the relative yield of pairs of charged and neutral mesons:
\beq
R^{c/n}={\sigma(e^+e^- \to P^+P^-) \over \sigma(e^+e^- \to P^0
{\bar P}^0)}~,
\label{rcn}
\eeq
where $P$ stands for the pseudoscalar meson, i.e. $B$, $D$, or $K$, and
dedicated measurements of such ratio have been done at the $\Upsilon(4S)$
resonance\,\cite{rups} at $\psi(3770)$\,\cite{rpsi} and at
$\phi(1020)$\,\cite{rphi}.

The values of the ratio $R^{c/n}$ at all three discussed resonances
are close to one due to these resonances being isotopic scalars, and
it is the deviation of the discussed ratio from one that presents
phenomenological interest. This deviation is generally contributed
by the following factors: the isospin violation due to the Coulomb
interaction between the charged mesons and due to the isotopic mass
difference between charged and neutral mesons, and, in the case of
the $K {\bar K}$ production at the $\phi(1020)$ resonance, a
non-negligible nonresonant isovector production amplitude. The
latter effect can be studied and described as the ``tail of the
$\rho$ resonance", while the isospin breaking due to the mass
difference is usually accounted for as a kinematical effect in the
$P$ wave production cross section factor $p^3$, where $p$ is the the
c.m. momentum of each of the mesons. The Coulomb effect has
attracted a considerable theoretical attention. The expression for
this effect in the ratio $R^{c/n}$ in the limit, where the resonance
and the charged mesons are considered as point-like
particles\,\cite{am} has the simple textbook form:
\beq \drcn = {\pi \alpha \over 2 v},
\label{drcn}
\eeq
with $\alpha$ being the QED constant and $v$ the velocity of each of the
(charged) mesons in the c.m. frame. However for the production of the real-life
mesons the analysis is complicated by the charge form factors of the
mesons\cite{lepage}, by the form factor in the vertex of interaction of the
resonance with the meson pair\,\cite{lepage,be} and generally by the strong
interaction between the mesons\,\cite{kmm,mv1,mv2}. In particular, it has been
argued\,\cite{mv1,mv2} that the modification of the Coulomb effect by the strong
(resonant) interaction between the mesons is quite significant. The previously
considered picture of the strong interaction was however somewhat unrealistic.
Namely, it has been assumed\,\cite{mv1,mv2} that the wave function in the $I=1$
state of the meson pair is vanishing at short but finite distances, which would
correspond to a singular behavior of the strong interaction at finite distances.
In this paper we derive the formulas for the Coulomb effect in the ratio
$R^{c/n}$ under the standard assumption about the strong scattering amplitude in
the channels with $I=0$ and $I=1$. We find that in the case of the
$\Upsilon(4S)$ and $\psi(3770)$ resonances, where the heavy meson pairs are
produced by the isotopically singlet electromagnetic current of the
corresponding heavy quark, the strong-interaction effect in the Coulomb
correction depends on the scattering phase $\delta_1$ in the $I=1$ channel and
is a smooth function of the energy across the resonance, while in the case of
the Kaon production at and near the $\phi(1020)$ there is also a smooth
dependence on the nonresonant part of the strong scattering phase $\delta_0$ in
the isoscalar channel inasmuch as there is a contribution of the isovector
production amplitude at these energies. In either case we find that the behavior
of the Coulomb effect is smooth on the scale of the resonance width, unlike the
behavior previously found\,\cite{mv1,mv2} under less realistic assumptions.

We further notice that essentially the same calculation can be
applied to considering the effect on the ratio $R^{c/n}$ of the
isotopic mass difference $\Delta m$ between the charged and neutral
mesons, at least in the first order in $\Delta m$, by considering
the mass difference as a perturbation by a (constant) potential. In
this way we find that the result coincides with the linear in
$\Delta m$ term in the ratio of the kinematical factors $p^3$ only
in the limit of vanishing strong scattering phase. Once the latter
phase is taken into account, there arises a correction whose
relative contribution is determined by the parameter $(p \, a)$ with
$a$ being the characteristic range of the strong interaction. We
therefore conclude that the conventionally used $p^3$ approximation
for this effect may be somewhat applicable to the $K {\bar K}$
production at the $\phi(1020)$ resonance, where $p \approx
120\,$MeV, but becomes quite questionable for the $D {\bar D}$
production at the $\psi(3770)$, where $p \approx 280\,$MeV.

The strong-scattering phase in the $P$-wave state of mesons produced in $e^+e^-$
annihilation near the threshold is proportional to $p^3$. We therefore expect
the discussed effects of the strong interaction in the ratio $R^{c/n}$ to
exhibit a measurable variation with energy. A measurement of this variation can
thus provide an information on the strong scattering phases, which is not
readily available by other means.

The material in the paper is organized as follows. In Sec.~2 we consider the
production of meson pairs by an isosinglet source and derive the formula for the
correction to $R^{c/n}$ due to a generic isospin-violating interaction potential
$V(r)$ viewed as a perturbation. In Sec.~3 we generalize this treatment to the
situation where the source is a coherent mixture of $I=0$ and $I=1$. The
specific expressions corresponding to the Coulomb interaction and the isotopic
mass difference are considered in Sec.~4. Sec.~5 contains phenomenological
estimates of the constraints on the parameters of the strong interaction between
heavy mesons based on the currently available data\,\cite{rups,rpsi} for $B
{\bar B}$ and $D {\bar D}$ production. Finally, in Sec.~6 we summarize our
results.

\section{General formulas for an isoscalar source}
We start with considering the behavior of the scattering wave functions of a
meson-antimeson pair in the limit of exact isotopic symmetry, i.e. neglecting
any Coulomb effects and the isotopic mass difference. We adopt the standard
picture (see e.g. in the textbook \cite{ll}), where the strong interaction is
confined within the range of distances $r < a$, so that beyond that range, at $r
> a$ the motion of the mesons is free. The two relevant independent solutions to
the Schr\"odinger equation at $r > a$ for the radial wave function in the $P$
wave are the free outgoing wave
\beq
f(pr)=\left ( 1 +{i \over pr} \right ) \, e^{ipr}
\label{fout}
\eeq
and its complex-conjugate, $f^*(pr)$, describing the incoming wave. A general
wave function of a pair of neutral mesons, $\phi_n(r)$ as well as of a pair of
charged mesons, $\phi_c(r)$, in this region is a linear superposition of these
two solutions.

In the region of strong interaction, i.e. at $r<a$, the isotopic symmetry
selects as independent channels the states with definite isospin, $I=0$ and
$I=1$, corresponding to the wave functions $\phi_0=\phi_c+\phi_n$ and
$\phi_1=\phi_c-\phi_n$. The detailed behavior of the $I=0$ and $I=1$ wave
functions inside the strong interaction region is not important for the present
treatment, and the important point is that the non-singular at $r=0$ `inner'
wave functions match at $r=a$ particular linear superpositions of the incoming
and outgoing waves (which superpositions in fact correspond to standing waves):
\begin{eqnarray}
&&\chi_0(r)= e^{i \delta_0} \, f(pr) +  e^{-i \delta_0} \, f^*(pr)\,, \nonumber
\\
&&\chi_1(r)= e^{i \delta_1} \, f(pr) +  e^{-i \delta_1} \, f^*(pr)\,,
\label{match}
\end{eqnarray}
where $\delta_0$ and $\delta_1$ are the strong scattering phases in respectively
the isoscalar and isovector states.

Consider now the production of meson pairs by a source  localized inside the
region of strong interaction, i.e. $r < a$, such as e.g. the electromagnetic
current. The wave function of the produced meson pairs at $r \le a$ is then
determined by both the source and the strong interaction, and the relevant
solution to the Schr\"odinger equation is chosen by the requirement that
asymptotically at large distances, $r \to \infty$, only an outgoing wave is
present. Let us first consider the simple case where the relevant
electromagnetic current is a pure isotopic singlet, which is the case for $D
{\bar D}$ and $B {\bar B}$ pair production. Then in the limit of exact isotopic
symmetry the outgoing waves for the `$n$' and the `$c$' channels have exactly
the same amplitude, which for our present purpose can be chosen as one:
\beq
\phi_c^{(0)}(r) = f(pr)~~{\rm and}~~ \phi_n^{(0)}(r) = f(pr)~~{\rm at}~r \to
\infty~,
\label{phi0}
\eeq
where the superscript $(0)$ stands for the approximation of exact isotopic
symmetry. It can be noted that the approximation of the free motion beyond the
region of the strong interaction in fact makes the expressions in
Eq.(\ref{phi0}) applicable at all $r > a$, i.e. all the way down to the matching
point $r=a$. It is helpful to notice for a later discussion that at the matching
point the $I=1$ wave function is vanishing while the $I=0$ function
$\phi_0^{(0)}$ contains only the outgoing wave. When continued into the strong
interaction region, i.e. at $r < a$, the function $\phi_0$ evolves into
the solution determined by the strong interaction and the source.

The isospin-violating effects of the Coulomb interaction and of the mass
difference $\Delta m$ between the charged and neutral mesons can be generally
described as being due to a presence of an extra potential $V(r)$ in the `$c$'
channel beyond the region of the strong interaction: $V=-\alpha/r$ for the
Coulomb interaction effect and a constant potential $V=2 \, \Delta m$ describing
the mass difference. In other words the wave function $\phi_n$ of the `n'
channel is still determined at $r > a$ by the radial Schr\"odinger equation for
free $P$-wave motion\footnote{Clearly, in the considered here first order in the
isospin violation only the difference of the interaction between the two
channels is important, thus any such difference can be relegated to one channel,
while keeping the other one unperturbed. Also, any effect of the mass difference
in the kinetic term $p^2/m$ is of order $v^2/c^2$ as compared to the discussed
here effect of $\Delta m$ in the overal energy difference between the two
channels, and is totally neglected in our treatment.}, while the equation for
the `$c$'
channel function $\phi_c$ reads as
\beq
\left ( {\partial^2 \over \partial r^2} + p^2 - m \,V(r) - {2 \over r^2} \right
) \, \phi_c(r)=0~.
\label{eqphicv}
\eeq

It is assumed throughout the present consideration that the isospin-breaking
potential exists only at distances beyond the range of the strong interaction,
i.e. that $V(r)$ has support only at $r > a$. The justification for such
treatment is that in the region of the strong force small isospin-violating
effects are compared to the energy of the strong interaction, so that the
contribution of any such effects arising at $r < a$ is very small, while in the
region $r > a$ the relative contribution of the potential $V(r)$ is determined
by its ratio to the kinetic energy of the mesons, which is small near the
threshold.

It should be emphasized that although the interaction at distances
$r > a$ is present only in the `$c$' channel, the wave functions in
{\it both} channels are modified in comparison with those in
Eq.(\ref{phi0}), as a result of the coupling between channels
imposed by the boundary conditions at $r=a$. According to the
setting of the problem of production of the meson pairs by a
localized source, the appropriate modified functions are those
containing at $r \to \infty$ only the outgoing waves
\beq
\phi_c \to
(1+x) \, f(pr), ~~~~ \phi_n \to (1+y) \, f(pr)~,
\label{phis}
\eeq
where the (complex) coefficients $x$ and $y$ arise due to the
potential $V$, and are proportional to $V$ in the considered here
first order of perturbation theory. These coefficients determine the
ratio of the production amplitudes: $A_c/A_n=1+x-y$, and the
discussed here modification of the yield ratio:
\beq
R^{c/n}=1+2\,{\rm Re}\, x - 2\,{\rm Re}\, y~.
\label{rcnxy}
\eeq

The modified wave function in both channels is subject to two conditions:\\
{\bf i:} The channel with neutral mesons has only an outgoing wave at all $r >
a$. In other words, the expression for $\phi_n(r)$ in Eq.(\ref{phis}) is valid
at all $r$ down to $r=a$;\\
{\bf ii:} The wave function of the channel with isospin $I=1$ at $r \le a$
should be
proportional to the standing-wave solution matching the function $\chi_1$ in
Eq.(\ref{match}), since there is no source for the $I=1$ state of the meson
pairs.\\
These two conditions are sufficient to fully determine the modified
functions at $r>a$ and thus to find the coefficients $x$ and $y$.

The first order in $V(r)$ perturbation of the wave function in the channel with
charged mesons is found in the standard way, using the P wave Green's
function $G_+(r,r')$ satisfying the equation
\beq
\left ( {\partial^2 \over \partial r^2} + p^2 - {2 \over r^2} \right ) \,
G_+(r,r') = \delta(r-r')~,
\label{gfe}
\eeq
and the condition that $G_+(r,r')$ contains only an outgoing wave when either of
its arguments goes to infinity. The Green's function is constructed from two
solutions of the homogeneous equation, i.e. from the functions $f(pr)$ and
$f^*(pr)$, as
\beq
G_+(r,r')={1 \over 2 \, i \, p} \left[ f(pr) \, f^*(pr') \, \theta(r-r') +
f(pr') \, f^*(pr) \, \theta(r'-r) \right ]~,
\label{gsol}
\eeq
where $\theta$ is the standard unit step function. The perturbation $\delta
\phi_c$ is then found as
\beq
\delta \phi_c(r) =m \, \int_a^\infty G_+(r,r') \, V(r') \, f(pr')
\, dr'~.
\label{chisol}
\eeq
One readily finds from this explicit form of the solution that $\delta \phi_c$
contains only the outgoing wave at asymptotic distances $r \to \infty$:
\beq
\delta \phi_c \left. \right |_{r \to \infty} = -{i \over 2 v}
\, f(pr) \int_a^\infty V(r') \, |f(p r')|^2 \, dr'~,
\label{chipi}
\eeq
so that the coefficient $x$ is purely imaginary:
\beq
x = -{i \over 2 v}
\, \int_a^\infty V(r') \, |f(p r')|^2 \, dr'
\label{xres}
\eeq
and gives no contribution to the ratio of the production rates $R^{c/n}$
described by Eq.(\ref{rcnxy})\footnote{It can be noticed that the integral in
Eq.(\ref{xres}) is divergent, which corresponds to the infrared-divergent
behavior of the perturbation for the phase of the wave function, logarithmic for
the Coulomb interaction and linear for a constant potential. This slight
technical difficulty can be readily resolved, for our present purposes, by
introducing an infrared regularizing factor $\exp(- \lambda \, r)$ in the
potential and setting $\lambda \to 0$ in the end result.}.

Consider now the matching of the wave functions at $r=a$. In this
region of $r$ one has $r < r'$ in the integral in Eq.(\ref{chisol})
so that the correction in the `$c$' channel has only an incoming
wave:
\beq
\delta \phi_c(r) \left. \right |_{r \to a}  = \eta \,
f^*(pr) \label{dchia} \eeq with \beq \eta = -{i \over 2 v}\,
\int_a^\infty V(r') \, \left [f(p r') \right]^2 \, dr'~.
\label{etaex}
\eeq
The wave functions $\phi_0 = \phi_c+\phi_n$ and
$\phi_1=\phi_c-\phi_n$ corresponding to the states with isospin
$I=0$ and $I=1$ are then found as
\beq
\phi_0 \left. \right |_{r \to
a}=2 f(pr) + y \, f(pr) + \eta \, f^*(pr)~~~{\rm and}~~~ \phi_1
\left. \right |_{r \to a}= \eta \, f^*(pr) - y \, f(pr)~.
\eeq
One
can now apply the condition {\bf ii} to determine the coefficient
$y$. Indeed, the condition for the wave function $\phi_1$ at $r \to
a$ to be proportional to $f^*(pr)+e^{2i\delta_1} \, f(pr)$ requires
$y$ to be given by
\beq y=-\eta \, e^{2 i \delta_1}~.
\label{resy}
\eeq
Upon substitution in Eq.(\ref{rcnxy}) this yields
\beq
R^{c/n}=1+{ 1 \over v} \, {\rm Im}\left [ e^{2i \delta_1} \,
\int_a^\infty e^{2ipr} \, \left ( 1+ {i \over p r} \right )^2 \,
V(r) \, dr \right ]~.
\label{rcnres0}
\eeq

\section{Mixed isoscalar and isovector source}
The formula (\ref{rcnres0}) gives the general expression for the
isospin-breaking effect in the considered yield ratio for the case where the
mesons are produced by an isoscalar source. The presented consideration can also
be extended to a situation where the source is a general coherent mixture of an
isoscalar and isovector. The specific isotopic composition of the source
determines the ratio of the coefficients of the amplitudes of the running
outgoing waves in the $I=1$ and $I=0$ channels at the matching point $r=a$,
which ratio we denote as $A_1/A_0$, thus defining $A_1$ and $A_0$ as the
production amplitudes in the respective channels (in the limit of exact isotopic
symmetry). In this situation the generalization of the expressions in
Eq.(\ref{phi0}) for radial wave functions in the `outer' region $r > a$ in the
zeroth order in the isospin violation can be written as
\beq
\phi_c^{(0)}(r) = (A_0+A_1) \, f(pr)~~{\rm and}~~ \phi_n^{(0)}(r) = (A_0- A_1)
\,f(pr)~.
\label{phi01}
\eeq
The isospin violation in the asymptotic form of these wave functions at $r \to
\infty$ can then be parametrized, similarly to Eq.(\ref{phis}), by complex
coefficients $x$ and $y$ as
\beq
\phi_c \to (A_0+A_1) \,(1+x) \, f(pr), ~~~~ \phi_n \to (A_0-A_1) \,(1+y) \,
f(pr)~,
\label{phis1}
\eeq
so that the yield ratio is found from
\beq
R^{c/n}=\left | {A_0+A_1 \over A_0-A_1} \right |^2 \left (1+2\,{\rm Re}\, x -
2\,{\rm Re}\, y~ \right ).
\label{rcnxy1}
\eeq

The coefficient $x$, similarly to the previous discussion and the equation
(\ref{xres}), is purely imaginary and in fact does not contribute in
Eq.(\ref{rcnxy1}), while the coefficient $y$ is found from the appropriately
modified conditions on the wave functions. Namely, the previously discussed
condition {\bf i} remains applicable, so that the asymptotic expression in
Eq.(\ref{phis1}) for the `$n$' channel function remains valid in the entire
`outer' region $r > a$ down to the matching point $r=a$. In order to allow for
the isovector component of the source the condition {\bf ii} has to be modified
as will be described few lines below.

The perturbation by the potential $V(r)$ of the `$c$' channel wave function at
the matching point $r=a$ is readily found, similarly to Eq.(\ref{dchia}), as
\beq
\delta \phi_c(r) \left. \right |\,_{r \to a}  = \eta \, (A_0+A_1) \, f^*(pr)
\label{dchia1}
\eeq
with $\eta$ given by Eq.(\ref{etaex}).

One can now write the expressions for the resulting `outer' wave functions in
the isotopic channels at the matching point:
\begin{eqnarray}
&&\phi_0(r) \left. \right |\,_{r \to a} = 2\, A_0 \, f(pr) + \eta \, (A_0+A_1)
\, f^*(pr)+ y \, (A_0-A_1) f(pr) = \nonumber \\
&&\left [ 2 \, A_0 + y \, (A_0-A_1) - \eta \, (A_0+A_1) \, e^{2 i \delta_0}
\right ] \, f(pr) +  \eta \, (A_0+A_1) \, e^{ i \delta_0} \, \chi_0(r)
\label{phi011}
\end{eqnarray}
and
\begin{eqnarray}
&&\phi_1(r) \left. \right |\,_{r \to a} = 2\, A_1 \, f(pr) + \eta \, (A_0+A_1)
\, f^*(pr)- y \, (A_0-A_1) f(pr) = \nonumber \\
&&\left [ 2 \, A_1 - y \, (A_0-A_1) - \eta \, (A_0+A_1) \, e^{2 i \delta_1}
\right ] \, f(pr) +  \eta \, (A_0+A_1) \, e^{ i \delta_1} \, \chi_1(r)~,
\label{phi111}
\end{eqnarray}
with  $\chi_0$ and $\chi_1$ being the standing wave functions from
Eq.(\ref{match}) in the corresponding isotopic channels, which when evolved in
the region of strong interaction contain no singularity at $r=0$. The remaining
parts in the latter expressions for the functions $\phi_0$ and $\phi_1$ describe
the proper {\it running} outgoing waves. These parts, when continued down in $r$
into the strong interaction region evolve to  match the
source at $r < a$. The ratio of the amplitudes of the isovector and the
isoscalar running waves is determined by the isotopic composition of the source,
and by the isotopically symmetric propagation through
the strong-interaction region. Thus the ratio of the amplitudes of these waves
at $r=a$ does not depend on the isospin-breaking effects at $r>a$ and should be
equal to $A_1/A_0$. Applying this condition to the isotopic wave functions given
by the expressions (\ref{phi011}) and (\ref{phi111}), one finds the equation for
the coefficient $y$:
\beq
{ 2 \, A_1 - y \, (A_0-A_1) - \eta \, (A_0+A_1) \, e^{2 i \delta_1} \over
2 \, A_0 + y \, (A_0-A_1) - \eta \, (A_0+A_1) \, e^{2 i \delta_0}} = {A_1 \over
A_0}~.
\label{eqny}
\eeq
This equation in fact replaces in this more general situation the previously
discussed condition  {\bf ii}, which condition and the ensuing result in
Eq.(\ref{resy}) are readily recovered in the limit $A_1/A_0=0$ from
Eq.(\ref{eqny}).

Considering that both $y$ and $\eta$ are of the first order in the potential
$V$, it is sufficient to use the linear expansion of the equation (\ref{eqny})
in $y$ and $\eta$, finding in this way the solution for $y$ in the form
\beq
y=-\eta {A_0 \, e^{2 i \delta_1} - A_1 \, e^{2 i \delta_0} \over A_0-A_1}~,
\label{resy1}
\eeq
and thus arriving at the final formula for the relative yield:
\beq
R^{c/n}=\left | {A_0+A_1 \over A_0 - A_1} \right |^2 \,
\left \{ 1 + { 1 \over v} \, {\rm Im}\left [ {{A_0 \, e^{2i \delta_1}- A_1 \,
e^{2i \delta_0}} \over {A_0 - A_1 }}  \,
 \int_a^\infty e^{2ipr} \,
\left ( 1+ {i \over p r} \right )^2 \, \, V(r) \, dr \, \right ] \right \}~.
\label{rcnres1}
\eeq
Given that $A_0=|A_0| \, e^{i \delta_0}$ and  $A_1=|A_1| \, e^{i \delta_1}$, the
amplitude-dependent factor in this formula can also be written in terms of the
real ratio $\rho=|A_1/A_0|$ as
\beq
{{A_0 \, e^{2i \delta_1}- A_1 \, e^{2i \delta_0}} \over {A_0 - A_1 }}= e^{2i
\delta_1} \, {1- \rho \, e^{i (\delta_0-\delta_1)} \over 1- \rho \, e^{-i
(\delta_0-\delta_1)}}~.
\label{arat}
\eeq

\section{The Coulomb and the mass-difference effects}

The general formulas in Eq.(\ref{rcnres0}) and (\ref{rcnres1}) can now be
applied to a discussion of the specific isospin-breaking effects in the $e^+e^-$
production of meson pairs at and near the threshold resonances. We start with
considering the effect of the Coulomb interaction. In a detailed treatment of
this correction one should include the realistic form factors of the mesons,
which cut off at short distances the difference in the electromagnetic
interactions between the charged and neutral mesons. In the present discussion
we replace for simplicity the gradual cutoff of the Coulomb interaction by an
abrupt cutoff at an effective range $r=a_c$, where generally $a_c \ge
a$\,\footnote{As previously mentioned, any extension of the isospin-breaking
potential inside the strong interaction region can result only in very small
corrections.}. The master integral with the Coulomb potential $V(r) = -
\alpha/r$ in the equations (\ref{rcnres0}) and (\ref{rcnres1}) then takes the
form
\begin{eqnarray}
&&\int_{a_c}^\infty e^{2ipr} \,
\left ( 1+ {i \over p r} \right )^2 \, \, V(r) \, dr = \nonumber \\
&&\alpha \left \{ \left [{\cos 2 pa_c \over 2
(pa_c)^2}+{\sin 2 pa_c \over pa_c}-Ci(2pa_c) \right ] + i \, \left[ {\pi \over
2}-{\cos 2 pa_c
\over pa_c} + {\sin 2 pa_c \over 2 (pa_c)^2} - Si(2pa_c) \right ] \right \} =
\nonumber \\
&& \alpha \left \{ \left [{1 \over 2 \, (p a_c)^2} - \ln (2 \, p a_c) +1 -
\gamma_E \right ] + i \, \left [ {\pi \over 2} + {pa_c \over 3} \right ] + O
\left [ (pa_c)^2 \right ] \right \}~,
\label{cints}
\end{eqnarray}
where the integral sine and cosine are defined in the standard way:
$$ Si(z)=\int_0^z \sin t \, {dt \over t}~~~~{\rm and}~~~~Ci(z)=-\int_z^\infty
\cos t \, {dt \over t}~,$$
and $\gamma_E = 0.577\ldots$ is the Euler's constant. The latter line in
Eq.(\ref{cints}) shows few first terms of the expansion of the integral in the
parameter $(p a_c)$. This expansion illustrates the behavior of the correction
toward the threshold. For the purpose of this illustration one can consider
first the simpler expression in Eq.(\ref{rcnres0}). The imaginary part, which
determines the discussed Coulomb effect in $R^{c/n}$ in the limit where there is
no strong scattering, $\delta_1 \to 0$, is not singular at $p a_c \to 0$, and
the textbook formula (\ref{drcn}) is recovered in this limit. The real part of
the integral in Eq.(\ref{cints}) is singular at small $p a_c$, but it multiplies
in Eq.(\ref{rcnres0}) the factor $\sin \delta_1$. The $P$-wave scattering phase
in its turn is proportional at small momenta to $p^3$: $\delta_1 \sim (pa)^3$,
so that the overall contribution of the real part of the integral is not
singular at the threshold either. Considering a more general expression for the
Coulomb effect for the case of an isotopically mixed source, following from the
equation (\ref{rcnres1}), one can readily arrive at the same conclusion that the
singular in $(p a_c)$ real part of the integral (\ref{cints}) does not lead to
an actual singularity, since it only enters the ratio $R^{c/n}$ multiplied by a
combination of the phases $\delta_0$ and $\delta_1$ (cf. Eq.(\ref{arat})), each
vanishing as $p^3$ toward the threshold.

As previously mentioned, the effect of the isotopic mass difference corresponds
to that of a constant potential $V = 2 \, \Delta m$ extending from the range of
the strong interaction $r=a$ to infinity. The master integral with such
potential  has the form
\begin{eqnarray}
&&\int_{a}^\infty e^{2ipr} \,
\left ( 1+ {i \over p r} \right )^2 \, \, V(r) \, dr = \nonumber \\
&&- {\Delta m \over p} \left \{ {2 \, \cos 2 \, p a \over p a}+ \sin 2 \, p a +
i \, \left [ {2 \sin 2 \, p a \over p a} - \cos 2 \, p a \right] \right \} =
\nonumber \\
&& - {\Delta m \over p} \left \{ {2 \over  p a} - 2 \, pa + 3 \, i + O \left [
(pa)^2 \right ] \right \}~.
\label{mints}
\end{eqnarray}
In the limit of vanishing strong scattering phases the mass
correction to $R^{c/n}$ is determined by only the imaginary part of
the integral, which in the limit of small $pa$ thus yields
\beq
R^{c/n}=1-{3 \, \Delta m \over v \, p} = 1-{3 \, \Delta m \over E}~,
\label{mrcnapp}
\eeq
where $E$ is the total kinetic energy of the meson pair, and the found
expression coincides with the linear in $\Delta m$ term in the expansion of the
usually assumed ratio of the kinematical factors $(p_+/p_0)^3$. Clearly, in the
more realistic case of presence of the strong scattering the real part of the
integral in Eq.(\ref{mints}) also contributes and the simple kinematical
approximation is generally invalidated.

\section{Phenomenological estimates}
In this section we discuss application of our formulas to interpreting the data
on the charged to neutral meson yield ratio $R^{c/n}$ at the near-threshold
resonances $\Upsilon(4S)$, $\psi(3770)$ and $\phi(1020)$. The purpose of this
discussion is to illustrate the effect of the strong scatering on the isospin
breaking corrections, and we use here the simplified picture of a abrupt cutoff
of the Coulomb interaction and of the isotopic mass difference effects. Such
simplification generally can be used as long as the parameter $(p a)$ is not
large. A detailed analysis should likely involve a model of a gradual cutoff,
since the details of the transition become important at lager momenta.

\subsection{$\Upsilon(4S)$}

The simplest case for the study of the isospin breaking corrections in the
relative production of heavy mesons is offered by the $B {\bar B}$ pair
production near and at the $\Upsilon(4S)$ resonance. Indeed, this process only
is due to the purely isosinglet electromagnetic current of the $b$ quarks, and
the isotopic mass difference between the $B$ mesons is very small: $\Delta m_B =
- 0.33 \pm 0.28\,$MeV\,\cite{pdg}, so that any deviation of the ratio $R^{c/n}$
from one is essentially entirely due to the Coulomb interaction. On the other
hand, the parameter $\alpha/v$ for the Coulomb effect in this case is the
largest due to small velocity of the $B$ mesons: at the energy of the
$\Upsilon(4S)$ peak $v_B/c \approx 0.06$. In particular, the numerical value in
the expression (\ref{drcn}) is $0.19$. The experimental data\,\cite{rups}
however indicate a significantly smaller deviation of $R^{c/n}$ from one. The
BaBar data with the smallest errors give $R^{c/n} = 1.006 \pm 0.036 \pm 0.031$.
Such behavior is likely a result of a combined effect of the meson and
production vertex form factors\,\cite{lepage,be} and of the discussed here
modification of the Coulomb correction by the strong scattering phase. These
effects can in principle be separated and studied quantitatively by measuring
the energy dependence of the ratio $R^{c/n}$ near the $\Upsilon(4S)$ resonance.
With the presently available data we can only use a simplified parametrization
of the form factor effects by introducing an abrupt cutoff for the Coulomb
interaction at $r=a_c \ge a$ and thereby estimate the likely regions in
the $(a_c, \delta_1)$ plane. Such estimate from the equations (\ref{rcnres0})
and (\ref{cints}) is shown in Fig.1 as a one-sigma area, corresponding to the
BaBar data with the statistical and systematic errors added in quadrature:
$R^{c/n}=1.006 \pm .048$. Clearly, more precise data from dedicated measurements
of the ratio $R^{c/n}$ are needed for a better understanding of the parameters
of strong interaction between the $B$ mesons.

\begin{figure}[ht]
  \begin{center}
    \leavevmode
    \epsfxsize=8cm
    \epsfbox{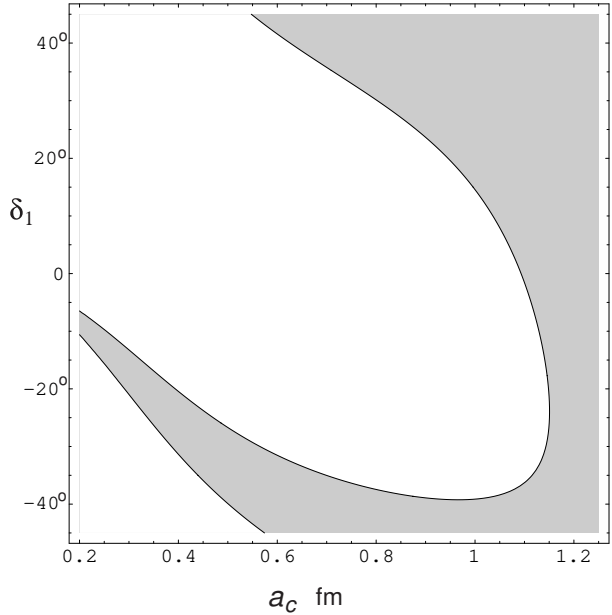}
    \caption{The one sigma area (shaded) in the $(a_c, \delta_1)$ plane
corresponding to the BaBar data on the $B^+B^-/B^0 {\bar B^0}$ yield ratio at
the $\Upsilon(4S)$ resonance.}
  \end{center}
\end{figure}

\subsection{$\psi(3770)$}
The largest isospin-breaking effect in the $D {\bar D}$ production
at the $\psi(3770)$ is that due to the mass difference between the
charged and the neutral $D$ mesons: $\Delta m_D = 4.78 \pm
0.10\,$MeV\,\cite{pdg}. The most precise measurements of this
process have been done\,\cite{rpsi} at the energy
$\sqrt{s}=3773\,$MeV. At this energy the momentum of each charged
$D$ meson is $p_+=254\,$MeV and that for a neutral $D$ meson is
$p_0=287\,$MeV. Thus the ratio of the kinematical factors
$(p_+/p_0)^3 \approx 0.69$ is significantly less than one. The
Coulomb effect is somewhat smaller. Indeed, the velocity of a
charged meson at this energy is $v_+/c=0.135$ and the expression
(\ref{drcn}) gives numerically $0.085$. One can notice that if  the
kinematical and the Coulomb factors are combined in a
straightforward way to estimate $R^{c/n} = (p_+/p_0)^3 \, [ 1 + \pi
\alpha/(2v_+)] \approx 0.75$, this would be in a very good agreement
with the experimental number\,\cite{rpsi}: $R^{c/n}=0.776 \pm
0.024^{+0.014}_{-0.006}$. Thus it is quite likely that at this
particular energy there is a considerable cancelation between the
strong-interaction effects in the yield ratio, and such cancelation
by itself imposes constraints on the parameters of strong
interaction between the $D$ mesons, which constraints is interesting
to analyze.

An analysis of the strong-interaction effects along the lines discussed in the
present paper generally runs into two difficulties. One is that our approach is
accurate only in the linear in $\Delta m$ approximation, while the actual effect
of the isotopic mass difference between the $D$ mesons is not very small.
However, numerically, the first term in the expansion of the kinematical factor
(Eq.(\ref{mrcnapp})) gives $0.67$, which is quite close to the mentioned above
value $0.69$, and it looks like the linear term gives a reasonable
approximation. The other point is that the cutoff parameter $a_c$ for the
Coulomb interaction at short distances does not necessarily coincide with the
range parameter $a$ used for the short-distance cutoff of the effect of the mass
difference. However, as previously noted, the Coulomb effect is somewhat small
at the energy of the $\psi(3770)$ resonance, and for the purpose of preliminary
estimates we set $a_c=a$ in our numerical analysis. In order to allow for
possible errors introduced by our approximations in comparing with the data, we
linearly add a theoretical uncertainty of $0.03$ units to the combined in
quadrature statistical and experimental errors. Proceeding in this way we find
that the only region in the $(a, \delta_1)$ plane at $a < 2\,$fm  consistent
with the CLEO-c data at one sigma level is the one shown in Fig.2.

\begin{figure}[ht]
  \begin{center}
    \leavevmode
    \epsfxsize=8cm
    \epsfbox{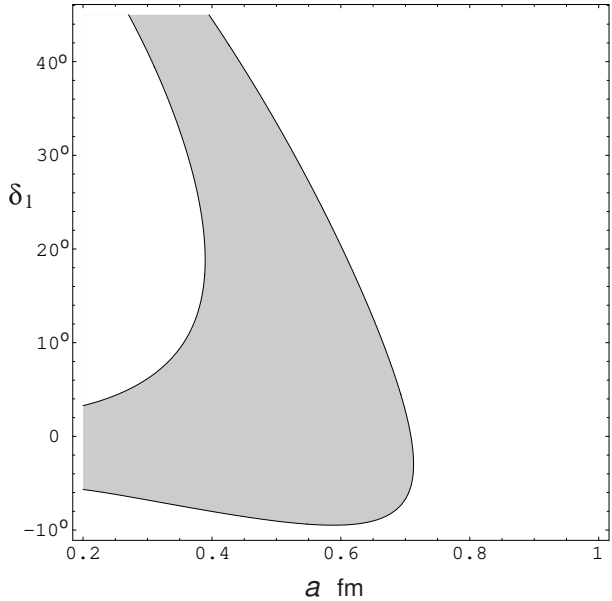}
    \caption{The area (shaded) in the $(a, \delta_1)$ plane corresponding to the
CLEO-c data on the $D^+D^-/D^0 {\bar D^0}$ yield ratio at the $\psi(3770)$
resonance. The uncertainty shown includes a one sigma experimental error with
our estimate of the theoretical uncertainty added linearly.}
  \end{center}
\end{figure}

It is interesting to compare the plots in the Figures 1 and 2. In the heavy
quark limit applied to both $b$ and $c$ quarks the strong interaction between
the heavy mesons should be the same, corresponding to the same range parameters
$a$ and $a_c$. The scattering phase $\delta_1$ for these two systems is
generally different due to different masses. However, provided there are no
isovector `molecular' bound states, the sign of the phase should be the same,
with the absolute value of the phase for heavier $B$ mesons being larger than
for the $D$ mesons. The comparison with the data for the $D$ mesons favors small
values of the range parameter, as indicated by Fig.2. If one also assumes that
$a_c \approx a$ for the $B$ mesons, the short range of $a_c$, according to
Fig.1, is compatible with the $B$ mesons data at a negative scattering phase
$\delta_1$, which sign of $\delta_1$ is also in agreement with the $D$ meson
data. A negative sign of $\delta_1$ corresponds to a repulsion, which for the
$I=1$ state of heavy meson pairs can be expected on general grounds\,\cite{ov}.

\subsection{$\phi(1020)$}
We believe that the production of $K {\bar K}$ pairs in $e^+e^-$
annihilation at and near the $\phi(1020)$ resonance merits a
separate analysis along the lines discussed in the present paper and
using detailed data similar to those in Ref.\cite{rphi}. As is
known, this production receives a small but measurable nonresonant
contribution from the isovector part of the electromagnetic current
of the $u$ and $d$ quarks, which corresponds to  an isotopically
mixed source. Furthermore, it has been pointed out\,\cite{belp} that a detailed
theoretical analysis of the $K^+K^-/K^0 {\bar K}^0$ yield ratio at the
$\phi(1020)$ resonance produces a result which possibly is at a meaningful
variance with the data.

At present we limit ourselves to noticing that the formula in
Eq.(\ref{rcnres1}), applicable in this situation, describes a smooth
behavior of the considered isospin breaking effects across the
resonance in the $I=0$ channel. Indeed, the $I=0$ scattering phase
at energy $E$ near the resonance energy $E_0$ is given by the
Breit-Wigner formula
\beq
e^{2 i \delta_0}= {\Delta - i \, \gamma
\over \Delta+i \, \gamma} \, e^{2 i {\tilde \delta}_0}~,
\label{bw}
\eeq
where $\Delta=E-E_0$, ${\tilde \delta}_0$ is the nonresonant
scattering phase in the isoscalar channel, and $\gamma$ is the width
parameter. Both ${\tilde \delta}_0$ and $\gamma$ are smooth
functions of the energy proportional to $p^3$ at small momentum, and
$\gamma(E_0)$ determines the resonance width $\Gamma$ as
$\gamma=\Gamma/2$. The ratio of the isovector and isoscalar
production amplitudes can then be parametrized near the resonance as
\beq
{A_1 \over A_0}= {\Delta+ i \, \gamma \over \mu} \, e^{ i \,
(\delta_1 - {\tilde \delta}_0)}~,
\label{a1a0rat}
\eeq
where $\mu$
is a parameter with dimension of energy: $\mu \sim m_\phi-m_\rho$.
The amplitude ratio entering the correction factor in
Eq.(\ref{rcnres1}) can then be written in the form
\beq
{{A_0 \,
e^{2i \delta_1}- A_1 \, e^{2i \delta_0}} \over {A_0 - A_1 }}= e^{2 i
\delta_1} \, {\mu - (\Delta - i \, \gamma) \, e^{-i
(\delta_1-{\tilde \delta}_0)} \over \mu - (\Delta + i \, \gamma) \,
e^{+ i (\delta_1-{\tilde \delta}_0)}}~,
\label{arat2}
\eeq
which
manifestly shows that this ratio is a pure phase factor of a complex
quantity slowly varying across the $\phi(1020)$ resonance.

\section{Summary}

We have considered the effects of the isospin breaking by the Coulomb
interaction and by the isotopic mass difference in the relative yield $R^{c/n}$
of pairs of charged and neutral mesons near threshold by a compact source, such
as in the production of heavy mesons in $e^+e^-$ annihilation. These effects are
modified by the strong interaction scattering phases. The general formula for a
situation where the source is an arbitrary coherent mixture of an isoscalar and
isovector is given by Eq.(\ref{rcnres1}). In particular, for a purely isoscalar
source, which is the case for the $e^+e^-$ annihilation into $D {\bar D}$ and $B
{\bar B}$ pairs the strong-interaction effect is determined by the scattering
phase $\delta_1$ in the $I=1$ channel (Eq.(\ref{rcnres0})). As a practical
matter we find that under the standard assumptions about the strong scattering
amplitudes in the near-threshold resonance region the ratio $R^{c/n}$ has a
smooth behavior with energy showing no abnormal rapid variation on the scale of
the resonance width. The energy dependence of this ratio is rather determined by
the non-resonant scattering scattering phase(s). In the $P$-wave the phase
$\delta_1$ is proportional to $p^3$, so that a measurement of the behavior ratio
$R^{c/n}$ with energy can provide information on this phase, which is not
readily accessible by other means. The behavior of the ratio $R^{c/n}$ at larger
energies away from the threshold also depends on the details of the onset of the
strong interaction between the heavy mesons at short distances and on the
behavior of their electromagnetic form factors, and a study of this behavior can
provide an insight into these properties of the heavy-light hadrons.

\section*{Acknowledgements}
The work of MBV is supported, in part, by the DOE grant DE-FG02-94ER40823.

\end{document}